# Programmable Jumping-Droplet Condensation

*Shan Gao*[1,4], *Jian Qu*[1], *Dehui Wang*[2], *Zhichun Liu*[3,*] and *Weigang Ma*[4,*]

[1]School of Energy and Power Engineering, Jiangsu University, Zhenjiang 212013, China

[2]Institute of Fundamental and Frontier Sciences, University of Electronic Science and Technology of China, Chengdu, 611731, China

[3]School of Energy and Power Engineering, Huazhong University of Science and Technology, Wuhan 430074, China

[4]Key Laboratory for Thermal Science and Power Engineering of Ministry of Education, Department of Engineering Mechanics, Tsinghua University, Beijing 100084, China

*Corresponding Emails:

zcliu@hust.edu.cn (Z. L.);

maweigang@tsinghua.edu.cn (W. M.)

**Graphic abstract**

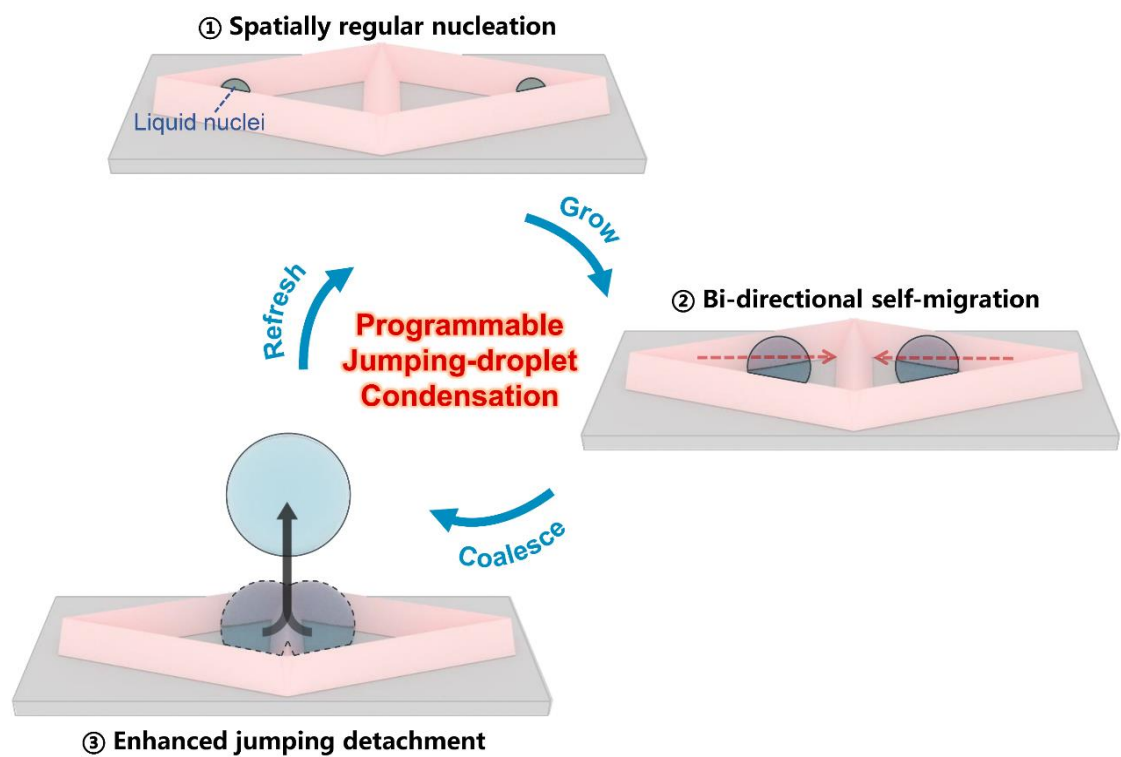

**Abstract**

Self-propelled droplet jumping during condensation has attractive prospects for energy harvesting, water collection and thermal management, but its real-life applications are greatly limited to the challenge of enabling a sustainable control on the entire droplet lifecycle. Herein, we propose a programmable jumping-droplet condensation that evolves along an artificially designed pathway without external stimulations, where the droplets can uniformly form at specific sites, spontaneously migrate and coalesce with their neighboring droplets, and jump off effectively to continuously refresh surface, significantly enhancing the heat transfer performance and durability of condensation. The programmable jumping-droplet condensation is achieved using a wedge-walled rhombus lattice structure surface inspired from the structures and functions of Namib desert beetle skin, shorebird beak and *setaria viridis* leaf vein. This surface integrates wetting contrast patterns with dual-gradient hierarchical structures, providing persistent and multidimensional droplet rectifications and thus realizing a sustainable control on the entire droplet lifecycle. Furthermore, we systematically investigate the morphology and behavior evolutions of droplets throughout their entire lifecycle, and fully elucidate the programmable control mechanisms of the lattice structure determined by its topology and wettability features. This work not only serves as theoretical foundations and reference framework to realize a durable jumping-droplet condensation and achieve its performance ceiling in a controlled manner, but also promotes the design and fabrication of functional structured surfaces for droplet manipulation and delivery, self-cleaning and anti-fogging/icing.

## 1. Introduction

Droplet manipulation has aroused great interests for its potential applications in diverse fields[1-3]. This fascinating phenomenon is not only employed in natural systems[4], such as cicadas[5], drain flies[6] and mushroom[7], serving to remove encapsulated particles, facilitate directional droplets transport and launch spores, but also has motivated researchers to explore its versatile applications in industrial contexts[8], including self-cleaning[9], microfluidics[10], water harvesting[11], electricity generation[12] and device cooling[13]. Well-controlled growing dew droplets are also essential to the heat and mass transfer performance of condensation[14], the most representative example is the coalescence-induced droplet jumping that brought a new breakthrough for conventional dropwise condensation[15]. This novel phase-change mode, termed "jumping-droplet condensation" could enhance heat transfer by up to 100%[16,17], and thus hold promise for solving the ultra-high heat flux dissipation in high-power electronics.

Therefore, the aspired goal of utilizing the jumping-droplet condensation in thermal management has been consistently pursued over the past decade[13,18,19]. However, there exists a significant bottleneck, namely the jumping mode of condensation would degrade into impaled flooding at larger supersaturations, and result in the deterioration of heat transfer[20]. To resolve this knotty issue, tremendous experimental efforts have been devoted to regulate the morphology of condensed droplet through specific surface structures[21-26], such as constraining droplet size[27] and dislodging droplet from surface structure gaps[28]. Furthermore, micro-nanoscale simulations were extensively performed to explore the initial nucleation and growth characteristics of condensates[29-33], that contribute to ascertaining the original cause of flooding and providing theoretical guidance for optimum design of surfaces[34,35].

Despite remarkable progress, it is still a challenge to implement the real-life applications of jumping-droplet condensation, probably because the exiting tactics have not achieved global regulation on droplet morphology evolution throughout the entire lifecycle. For the natural jumping-droplet condensation, chaotic nucleation sites generate the droplets trapped within substrate interstices, these free-growing droplets with various sizes and states touch their neighbors randomly, such a passive coalescence mode would seriously influence the jumping probability and efficiency[36,37]. Therefore, it necessitates comprehensive manipulations on the condensed droplets from nucleation to growth, motion and removal, enabling a programmable jumping-droplet condensation that can evolve along a desired pathway.

To address the aforementioned challenges, we proposed an innovative strategy without external stimulation to realize a programmable jumping-droplet condensation with an excellent heat transfer efficiency and durability, based on a multi-bioinspired surface that integrates dual-gradient hierarchical structures and wetting patterns. This condensation mode could enable full-lifecycle control on droplet evolutions, by rationally regulating nucleation sites, actuating condensates to move and actively coalesce with their neighbors, and persistently facilitating droplet jumping detachments. Combined with molecular dynamics (MD) simulations and theoretical analyses, the morphologic evolutions and spontaneous behaviors of condensed droplets were investigated systematically, to explore the sustainable and programmable control mechanisms of the newly designed surface on the dynamics of condensed droplets over its entire lifetime. The findings gained in this work can provide guidance for future efforts in achieving droplet manipulation, self-cleaning, water collection and heat transfer enhancement.

## 2. Results and Discussion

### *2.1. Multi-bioinspired and hierarchical structure surface*

To enhance the efficiency and durability of the jumping-droplet condensation, a hierarchical structure named wedge-walled rhombus lattice structure (WRLS) surface was proposed, and its design concept fuses unique mechanisms of three kinds of representative biological organism. As depicted in Figure 1, the shorebird beak can deliver liquid directionally during feeding behaviors[38], whose configuration is modeled with mirror symmetry to construct the secondary gradient structure, namely the rhombus lattice. The frame architectures of this lattice consist of an array of wedge-shaped ridges mimicking the main vein of *setaria viridis* leaf[39], which act as the primary gradient structure. Additionally, inspired by the wetting contrast patterns found on Namib desert beetle's skin for moisture capture[40], the converging sides of the hydrophobic rhombus lattices are modified with uniform hydrophilic spots. The detailed geometry and wettability of surface structure are illustrated in Section S2 of the Supplementary Information. The rhombus lattice with an opening angle $\alpha = 25°$ consists of the wedge-shaped ridges characterized by a width $W = 3.6$ nm, height $H = 1.25W$ and length $L = 6.74W$, and the water contact angles (WCA) for the hydrophobic structures and hydrophilic patterns are 126.5° and 87.8°, respectively.

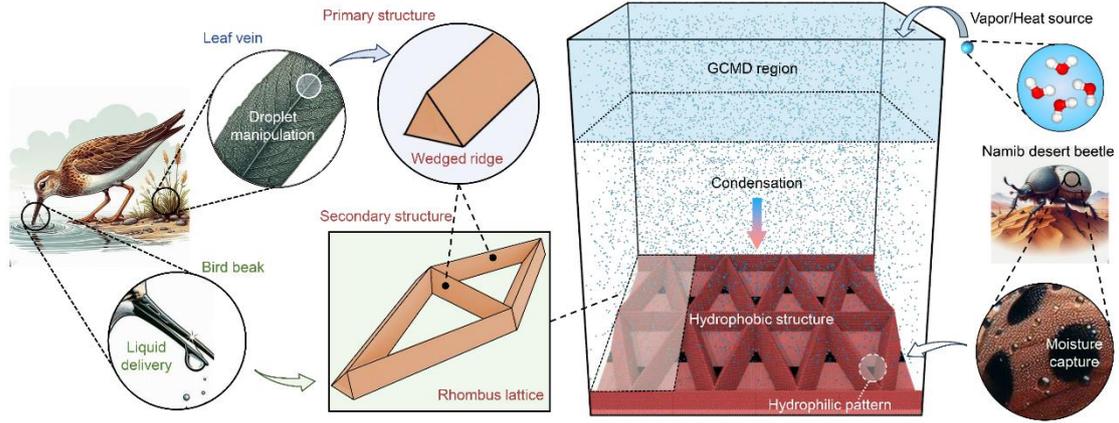

Figure 1. Design principle of the wedge-walled rhombus lattice structure (WRLS) for active jumping-droplet condensation. The wedge-shaped ridges construct an array of the water-repellent rhombus lattice structures that serve as the droplet rectifier, and hydrophilic patterns are uniformly embedded in the converging sides of the lattices to regulate nucleation sites.

## 2.2. Active jumping-droplet condensation via droplet rectifications

In the model system with dimensions of 47.9 nm × 54.3 nm × 54.3 nm, the cooling WRLS was initially submerged in the saturated water vapor to condense it into droplets, and for convenience, one single rhombus lattice structure was chosen to demonstrate the dynamic behaviors of condensates throughout their entire lifecycle. As shown in Figure 2, the hydrophilic spots within each lattice act like seeds to produce a deterministic and regular nucleation, that is, the liquid nuclei prefer to form at the opposite vertexes of the rhombus lattice. These condensed droplets inflate and migrate towards each other simultaneously during the growth stage, resulting in an in-plane rectification effect, moreover, their sizes remain roughly uniform under the shape restriction imposed by the lattice frame. When the growing droplets move to the lattice center and reach a sufficient size, they coalesce with each other and jump off efficiently, aided by the centrally located ridge, that activates an intensified out-of-plane rectification effect. Consequently, the substrate reverts to vacant lattices and initiate a new cycle of nucleation, growth, migration and detachment. This strategy propels droplets to transport, coalesce and jump actively without external energy input, by virtue of the multidimensional droplet rectifications, the resultant active coalescence-induced detachment events enable an efficient and consistent refresh of condensed droplets, that could improve the heat transfer efficiency and durability of jumping-droplet condensation.

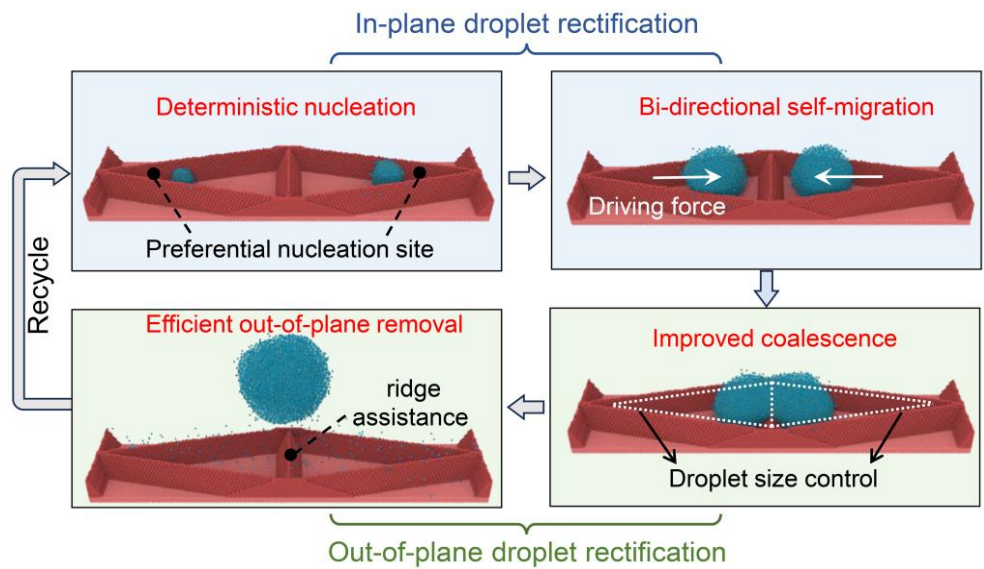

Figure 2. Programmable jumping-droplet condensation achieved by the persistent and multidimensional condensates rectifications.

Based on the entire simulated condensation process, we investigated and discussed the morphological evolution characteristics of droplets in the sequence of the droplet lifecycle. Furthermore, to analyze the sustainable and programmable control mechanisms for droplets imposed by the WRLS surface, another conventional structured surface, namely the rectangular-walled square lattice structure (RSLS) surface without wetting contrast patterns was employed as a reference group. Its projected area and the number of lattice structures, geometric dimensions and wettability are consistent with that of WRLS surface, with details given in Section S2 of the Supporting Information.

*2.3. Spatially regular and uniformly sized nucleation*

Figure 3a depicts the nucleation scenarios on the WRLS surface and RSLS surface at the same moment (5 ns), respectively. The initially dispersed vapor molecules aggregate into liquid nuclei and attach to the low-temperature surface structures, accompanied by the energy exchange, but strikingly, these nucleating embryos exhibit distinct spatial distributions. For the RSLS surface, while nucleation preferentially occurs within square lattices, the size, orientation and location properties of nucleating embryos appear quite chaotic. In contrast, on the WRLS surface, the nucleation sites are regularly situated in the converging sides of rhombus lattices, exhibiting a more regular nucleation pattern. To further quantitatively characterize the nucleation control capabilities of these two structured surfaces, we performed three sets of independent MD simulations, and their results are compared from two perspectives: spatial ordering (Figure 3b, c) and size uniformity (Figure 3d, e) of liquid nuclei. Additionally, we also studied the nucleation on uniformly hydrophobic WRLS surface, please see details in Section S5 of Supplementary Information.

First, the liquid nuclei coverage ratio $\varphi_c$, is calculated as the proportion of lattice structures undergoing nucleation, as illustrated in Figure 3b. There exist many vacant lattices on the RSLS surface throughout the nucleation incubation period ($\varphi_c = 75.0 \pm 5.0$ %), while almost every lattice of the WRLS surface contains the liquid nuclei ($\varphi_c = 97.1 \pm 3.6$ %), indicating a higher density and a spatially regular distribution of nucleation sites. Besides, the liquid nuclei at different locations are classified and counted to calculate the nucleation position proportion $\varphi_p$ for evaluating the orientation degree of nucleation sites, as shown in Figure 3c. For the RSLS surface, the liquid nuclei are randomly distributed at the four corners of the square lattices, with the same probability ($\varphi_{p1} = 26.1 \pm 6.7$ %, $\varphi_{p2} = 23.9 \pm 3.5$ %, $\varphi_{p3} = 25.9 \pm 5.6$ %, $\varphi_{p4} = 24.0 \pm 4.6$ %). By contrast, all of the liquid nuclei are regularly located in the converging sides of rhombus lattices on WRLS surface ($\varphi_p = 100$ %), revealing a superior orientation degree of the nucleation sites. In general, the WRLS surface possesses a more excellent control on the spatial ordering of liquid nuclei than that of conventional structured surfaces, and this regulation mechanism will be further discussed later by analyzing the nucleation energy barrier.

To quantitatively assess the size uniformity of the liquid nuclei, we extracted all the condensates at the initial nucleation moment (5 ns) and calculated their equivalent diameters, then divided them based on a size interval of 0.5 nm and plotted the size distribution curves, as shown in Figure 3d. Obviously, the nucleating embryos on the RSLS surface are scattered across a wide range of 1.25 nm to 4.75 nm, for the WRLS surface, however, the nuclei is predominantly concentrated within the narrow range of 3.75 nm to 4.75 nm, declaring an improved size uniformity. Furthermore, the median values $D$ and the standard deviations of nuclei diameters were also calculated, as depicted in Figure 3e. Due to the rapid and nearly synchronous nucleation, the resultant liquid nuclei on the WRLS surface show larger median values and smaller standard deviations of their equivalent diameters ($D_1 = 4.4 \pm 0.36$ nm, $D_2 = 4.2 \pm 0.42$ nm, $D_3 = 4.2 \pm 0.37$ nm), indicating a good control over the liquid nuclei size.

The nucleation regulation mechanism of the WRLS surface was analyzed according to the classical nucleation theory (CNT) [41]. For heterogeneous condensation, in order to form a stable liquid embryo with the critical nucleus radius $r_e$, the Gibbs free energy barrier of nucleation $\Delta G$ needs to be overcome, that can be expressed by[42]

$$\Delta G = -\frac{\rho V}{M} k_B N_A T_s \cdot \ln s + \sigma_{LV}(A_{LV} - A_{SL} \cdot \cos\theta) \qquad (1)$$

where $\rho$, $V$ and $M$ are the liquid density, the volume of embryo and the molar mass of liquid. $k_B$, $N_A$ and $T_s$ represent the Boltzmann constant, the Avogadro number and the surface temperature. $s$ is the supersaturation defined as the ratio of vapor pressure $P_\infty$ to the saturated vapor pressure $P_s$ at $T_s$. $\sigma_{LV}$, $A_{LV}$, $A_{SL}$ and $\theta$ represent the surface tension of liquid, the liquid-vapor interfacial area, the solid-liquid interfacial area and the Young's contact angle. Obviously, the energy barrier $\Delta G$ reflecting the nucleation difficulty is directly correlated with the liquid nucleus shape that affected by the topology and wettability features of solid surface. As illustrated in insets of Figure 3f, there are two typical nucleation configurations on the WRLS surface, namely the spherical-wedge nucleus respectively located at the hydrophilic converging side and the hydrophobic diverging side of the rhombus lattice, and their nucleation energy barriers $\Delta G_1$ and $\Delta G_2$ were calculated respectively, while for the RSLS surface, the nucleation energy barrier $\Delta G_3$ at the corner of the hydrophobic square lattice was also obtained based on the CNT model, please see calculation details in Section S4 of Supplementary Information. Additionally, we considered a classic nucleation configuration involving the energy barrier $\Delta G_0$, where a spherical-cap nucleus lies on a smooth hydrophobic surface. To compare the nucleation difficulty directly, the energy barriers $\Delta G_1$, $\Delta G_2$ and $\Delta G_3$ were normalized by $\Delta G_0$, as shown in Figure 3f. The lattice structures reduce the liquid-vapor boundary of nucleus and further lower the energy barrier to overcome, thus nucleation is more readily to occur in the structure intervals of RSLS surface, and follows the rules $\Delta G_1 < \Delta G_2 < \Delta G_3 < \Delta G_0$. More notably, nucleation is also affected by the topology feature (the opening angle $\alpha$) and the wettability of rhombus lattice. For a homogeneously hydrophobic rhombus lattice, it is revealed that $\Delta G_1$ at the converging side is already smaller than $\Delta G_2$ at the diverging side when the opening angle $\alpha < 60°$ (Figure S9 in Supplementary Information), this difference is further intensified by introducing the hydrophilic pattern into the converging side (Figure 3f). To summarize, the rhombus lattice structure greatly promotes the droplets nucleation compared to the conventional square lattice. Moreover, attributed to the lower nucleation energy barrier controlled by the opening angle $\alpha$ and the wettability of rhombus lattice, the vapor prefers to generate liquid embryo at the hydrophilic converging side, resulting in a spatially regular nucleation.

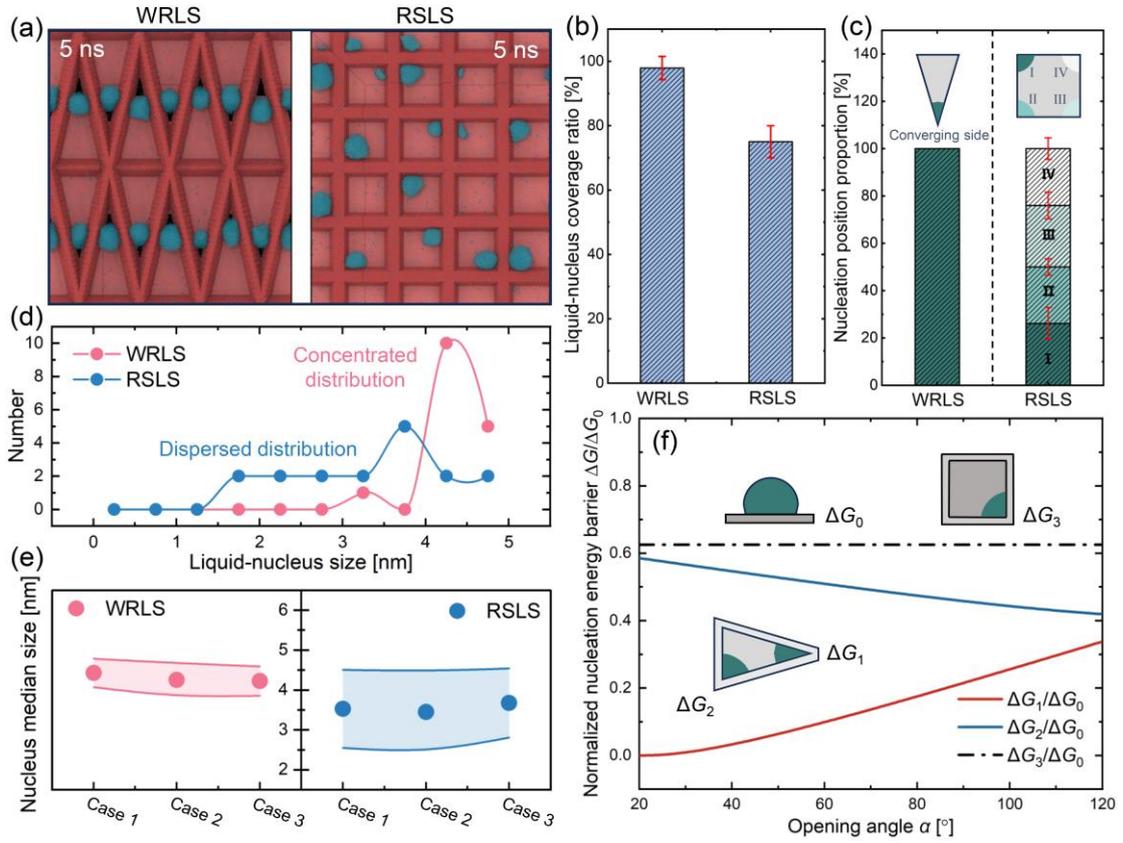

Figure 3. Spatially regular and uniformly sized nucleation: (a) snapshots of the nucleation scenarios on the WRLS surface and RSLS surface. Comparison for the spatial ordering of nucleating embryos between the WRLS surface and the RSLS surface: (b) liquid nuclei coverage ratio $\varphi_c$ defined as the proportion of lattice structures undergoing nucleation; (c) nucleation position proportion $\varphi_p$ evaluating the orientation degree of nucleation sites. Comparison for the size uniformity of nucleating embryos between the WRLS surface and the RSLS surface: (d) size distributions of the liquid nuclei; (e) median values of the nuclei diameters with standard deviations denoted by the shaded areas. Nucleation regulation mechanism of the WRLS surface: (f) normalized nucleation energy barriers at different positions and their dependencies on the opening angle $\alpha$ of rhombus lattice.

*2.4. Bi-directional self-migration accompanied with growth*

After preferentially nucleating at the converging sides, the condensed droplets grow inside the unit cell of rhombus lattices and move toward the central diverging sides simultaneously, as demonstrated in Figure 4a. To capture the droplet growth dynamics, the number of water molecules was counted to convert into droplet volume $V$ and radius $R$, assuming a spherical shape. Figure 4b records the temporal variations of the droplet radius, both the droplet radii increase linearly over time with steady and comparable growth rates after the initial dramatic inflation, $dR_1/dt = 1.31$ m/s and $dR_2/dt = 1.53$ m/s. Correspondingly, the increase of droplet volume $V$ versus time $t$ follows a power-law relation (Figure S11 in Supplementary Information). To investigate the characteristics of the droplet directional transport, the position coordinates of droplet centroid were extracted from the droplet trajectory, their temporal evolutions could be divided into two phases, as shown in Figure 4c. In the early self-migration period, both the droplet positions change steadily but exhibit obviously opposite trends, indicating an aggregated motion from lattice ends (converging side) to lattice center (diverging side), and the corresponding transport velocities were calculated to be $U_1 = 1.25$ m/s and $U_2 = 1.20$ m/s, respectively. Subsequently, when the liquid-vapor boundary of the droplet contacts the wedge-shaped ridge at the lattice center, the droplet positions turn to remain constant under the spatial confinement from lattice frame, until the droplet grows into a sufficient size and merge with its neighbors.

To probe the mechanisms of the spontaneous migration of the droplet from an energetic standpoint, we performed another set of simulation to calculate the internal energy variations, please see Section S8 in Supplementary Information for simulation details. Differing from the natural formation of condensate, a constant-sized droplet was constructed with its base was embedded in the converging end of the identical rhombus lattice artificially. As illustrated in insets of Figure 4d, to obtain a wetting state under equilibrium conditions and prevent the premature self-transport, the droplet was restrained around the original position with its center of mass could move freely in the vertical direction, by attaching a spring force. Then this constraint was removed to initiate spontaneous migration until the moving droplet stopped at a certain location. Figure 4d shows the temporal variations for the potential energy $E_\mathrm{p}$ of droplet and the interaction energy $E_\mathrm{i}$ between droplet and substrate during this process. At first, the potential energy $E_\mathrm{p}$ fluctuates within a higher value range, suggesting the equilibrium system is energetically metastable during the initial wetting period. However, shortly after the release of the constraint, the droplet potential energy decreases sharply and converts into the kinetic energy, propelling the droplet to move. Eventually, $E_\mathrm{p}$ stops falling and tend to fluctuate within a lower value range, indicating that the whole system has become energetically stable. This variation trend is also observed in $E_\mathrm{i}$, which is negative and decreases continuously during the directional transport, this is because the droplet seems like to be immersed in the lattice groove when it moves to the wider end, solid-liquid contact area increases accordingly and more liquid molecules get closer to the substrate, as a result, the droplet-substrate interactions is intensified. The above results indicate the directional self-migration behavior follows the principle of minimum energy, and the droplet at the diverging side is thermodynamically stable. In addition, it should be stressed that the droplet undergoes acceleration and deceleration stages successively, as shown in Figure S14c in Supplementary Information, due to its time-varying morphology and deformation, please refer to the detailed discussions in Section S8 of Supplementary Information.

To further provide insights into the mechanism governing directional transport from a hydrodynamics standpoint, we theoretically analyzed the physical forces acting on the growing droplet and investigated the dependency of transport velocity on the lattice structure topology. As schematically shown in Figure 4e, the droplet with a sectional length of $L$ is entrapped in the unit cell of the rhombus lattice with an opening angle $\alpha$, the lateral boundaries of droplet are restricted by the lattice frame walls and $X$ represents the distance between droplet tail and lattice tip. The inflation along the lattice groove causes the squeezed droplet to become fore-aft asymmetric. The curvature radius at the rear liquid−vapor interface $R_1$ is inconsistent with that of the front liquid−vapor interface $R_2$, resulting in a Laplace pressure difference $\Delta P$[43-46], which provides the driving force $F_L$ to overcome the viscous resistance $F_V$ and interfacial friction $F_f$, then facilitates the droplet to move downward the wider end of the lattice groove[47,48]. Considering the condensed droplet grows and moves in a state of dynamic equilibrium, thus it is reasonable that directional migration is regarded as a uniform linear motion, which is also corroborated in the present MD simulations. It follows that the droplet transport velocity $U$ could be determined by the force balance condition $F_L = F_V + F_f$, and is expressed as

$$U = 4\sigma\alpha X\tan\frac{\alpha}{2}(X + \frac{L}{2})(\frac{1}{R_1} - \frac{1}{R_2})/kL\eta \tag{2}$$

where $\sigma$ and $\eta$ are the surface tension and the dynamics viscosity of droplet, and $k$ denotes a coefficient (please see fully detailed derivation in Section S7 of Supplementary Information). It can be found that the transport velocity $U$ is dependent on both the length ratio $L/X$ characterizing the droplet deformation degree and the opening angle $\alpha$ representing the lattice structure topology (Figure S13 in Supplementary Information). Therefore, there are two ways to facilitate the self-migration of a condensed droplet in the rhombus lattice, one is to maintain a relatively small opening angle of lattice structure and the other is to increase the sectional length of droplet, namely enhancing the droplet deformation degree.

Besides, additional MD simulations were performed to verify the theoretically calculated result of the transport velocity. We chose the unit cell of the rhombus lattice with various opening angles $\alpha$, and conducted condensation simulations under the identical thermodynamic conditions, to analyze the self-migration of one single condensed droplet, see details in the Section S9 of Supplementary Information. The simulated transport velocity was calculated based on the spatial-temporal trajectory of droplet. Moreover, the length ratio $L/X$ was also extracted from the droplet morphology, it fluctuates slightly and stays almost constant during the growth of the condensed droplet (Figure S15b in Supplementary Information), which testifies the above theoretical viewpoint that the migration speed of a naturally growing droplet is approximately uniform. Then the theoretical value of the transport velocity was obtained by substituting the time-averaged value of $L/X$ into Equation 2, and Figure 4f shows the dependencies of the self-transport velocity $U$ with the opening angle $\alpha$ of rhombus lattice by the present theory and MD results. The theoretically predicted values of $U$ basically agree with the corresponding simulation values, and both of them decrease as $\alpha$ increases.

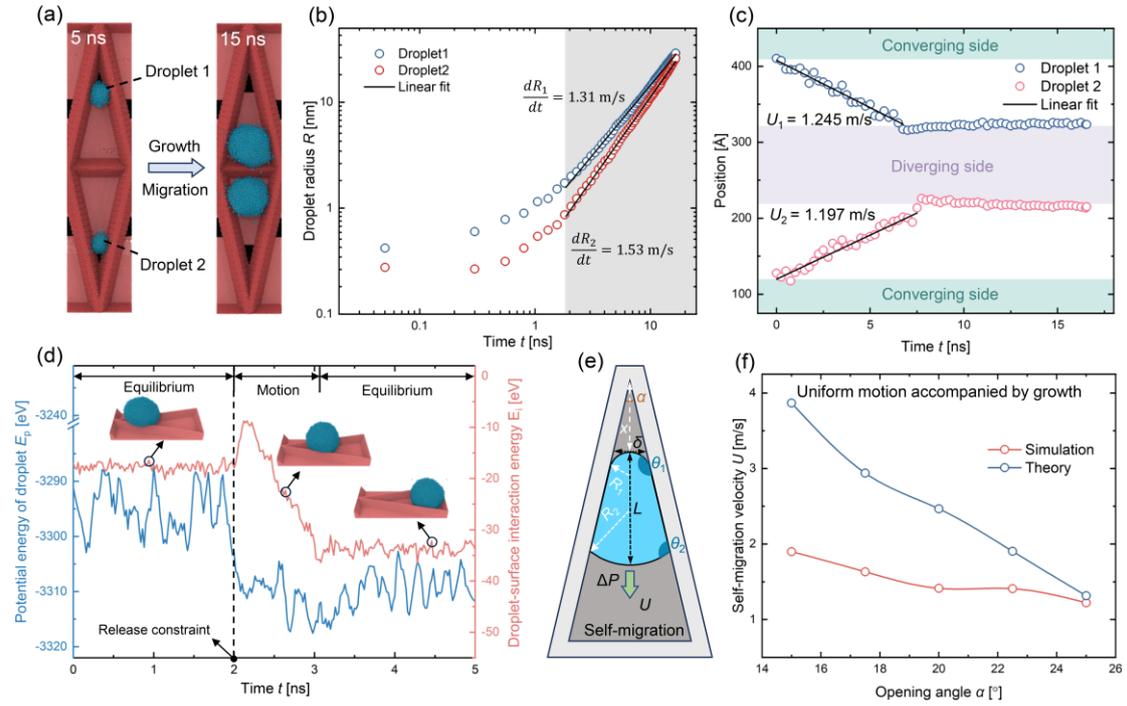

Figure 4. Bi-directional self-migration of the growing droplet: (a) time series snapshots of a unit cell of rhombus lattices show the droplets transporting spontaneously while growing; (b) Evolutions of the growing droplet radius $R$; (c) center-of-mass positions of the droplet originally located at the converging sides as functions of time. Energetic analyses of the droplet self-migration in the unit cell of rhombus lattice: (d) temporal variations of the droplet potential energy $E_p$ and the droplet-substrate interaction energy $E_i$ during the directional transport of a constant-sized droplets. Force analyses of the droplet self-migration in the unit cell of rhombus lattice: (e) schematic for the directional transport of a growing droplet driven by the asymmetric Laplace pressure; (f) variations of the transport velocity $U$ with the opening angle $\alpha$ of rhombus lattice based on the present theoretical predictions and MD simulations.

*2.5. Enhanced coalescence-induced jumping*

Following the bi-directional transport, the two adjacent droplets merge with each other at the lattice center, thereby leading to an enhanced out-of-plane removal from WRLS surface. Figure 5a shows the situations of coalescence for condensed droplets on WRLS surface and RSLS surface, respectively. It can be observed that the WRLS surface promotes the jumping detachments of droplets upon coalesce significantly, with the assistance of intermediate wedged ridge. Conversely, coalescence on the RSLS surface generates large residual droplets in a Cassie − Baxter state, covering the substrate surface and hindering its refreshment. To quantitatively assess and compare the capability of these two surfaces to enable droplet jumping, we recorded the size ratio *r* of large droplet to small droplet before coalescence and the outcome of each coalescence event over the same period from the initial nucleation to the end of the first refresh cycle of the WRLS surface, as shown in Figure 5b. Coalescence events occur continuously throughout the recording period for the RSLS surface and predominantly lead to jumping failures. However, coalescence events on the WRLS surface are regularly concentrated within a short time window, and almost all of them trigger the droplet jumping successfully, except the only one jumping failure caused by the occasional multi-droplet coalescence. Furthermore, the droplet size ratio before coalescence *r* shows an apparent discrepancy, its average value approaches to 1 more significantly on the WRLS surface, indicating the two droplets before coalescence could be approximately considered as identical. To sum up, the enhancement of coalescence-induced jumping by the rhombus lattice can be attributed to two aspects:

one is prompting the two-droplets coalescence events via the in-plane rectifications rather than the multi-droplets coalescence originated from the spatially regular nucleation, and the other is controlling the droplet size before coalescence to be approximately equal via the droplet shape restriction imposed by the lattice frame, both the factors have positive effects on the energy conversion efficiency of coalescence.

In addition, the wedge-shaped ridge located at the lattice center also plays a crucial role in enhancing droplet jumping. As shown in Figure 5c, the coalescence between two equal-sized droplets on a pillar array and a wedged ridge were investigated respectively, please see simulation details in Section S10 of the Supplementary Information. At the initial phase of coalescence on the pillar array surface, liquid mass from two droplets flows in reversed in-plane directions and collides into each other, causing an unnecessary energy dissipation. Whereas on the surface with a wedged ridge, the in-plane momentums are effectively diverted to the out-of-plane direction by the ridge, contributing to the promotion of jumping. The results in Figure 5c confirms that the jumping velocity $U_j$ decreases with the increasing droplet radius $R$, and $U_j$ on the wedged ridge are generally higher than that on the pillar array, with an about 5 to 10 times increasement. Moreover, we also simulated the coalescence between two unequal-sized droplets with radii $R_1$ and $R_2$ on a pillar array and a wedged ridge, respectively. As illustrated in insets of Figure 5d, on the surface with a wedge-shaped ridge, the difference in droplet radius creates a pressure gradient between the left and right side of the coalescing droplet, that makes the liquid mass in smaller droplet flow to the center of coalescing droplet more rapidly. As a result, the smaller droplet rolls toward the side of the larger droplet, using the wedged ridge as a pivot, that leads to a non-vertical jumping eventually.

Figure 5d demonstrates the variations of energy conversion efficiency $\eta$ with size mismatch $M$ of coalescing droplet. $\eta$ is defined as ratio of translational kinetic energy of jumping droplets $E_{k,tr}$ to the surface energy released upon coalescence $\Delta E_s$, namely, $\eta = E_{k,tr}/\Delta E_s \approx mU_j^2/8\pi\sigma \left[R_1^2 + R_2^2 - (R_1^3 + R_2^3)^{\frac{2}{3}}\right]$, where $m$ is the mass of coalesced droplet, $U_j$ is the jumping velocity and $\sigma$ is the surface tension of droplet. The coalescence mismatch $M$ is defined as the ratio of the radius difference to the average radius, $M = 2|R_1 - R_2|/(R_1 + R_2)$, that quantifies the size deviation of droplets before coalescence. Here, the radius of smaller droplet $R_1$ was kept fixed to be 8.1 nm, and the radius of bigger droplet $R_2$ was varied for different mismatches. It is found that the energy conversion efficiency descends as the size mismatch increases whether for the wedged ridge or the pillar array. But remarkably, on the surface with a wedged ridge, $\eta$ is improved by an order of magnitude compared to the pillar array, and the coalescence-induced jumping can be realized until a large size mismatch of 54.2% (corresponding to a size ratio 1:1.74), this cut-off mismatch is significantly higher than that of 22.0% (corresponding to a size ratio 1:1.25) on the pillar array. The results indicate that the wedged ridge enhances the energy conversion efficiency and increases the tolerance of the coalescence mismatch, and thus synergistically improving the probability of droplet jumping.

To further understand the mechanism of this coalescence-induced jumping enhancement, we calculated the variations of droplet-substrate interaction energy $E_i$ on the wedged ridge and the pillar array, respectively, as illustrated in Figure 5e. Based on the temporal evolution of $E_i$, the entire coalescence process could be roughly divided into three phases. Starting from the expansion of the liquid bridge connecting the two droplets, more liquid molecules move closer to substrate and intensify the droplet-substrate interaction force. Therefore, the value of $E_i$ decreases continuously until reaching a minimum value when the downward-moving liquid bridge impacts the substrate and arrives at its maximum spreading state (stage I). Subsequently, the coalesced droplet accelerates upwards under the reaction force exerted by the substrate, the droplet-substrate interaction recedes correspondingly as more liquid molecules move farther from substrate, thus the value of $E_i$ turns to increase and gradually reach its maximum value during the stage II. At last, the value of $E_i$ becomes steady at 0 eV, declaring the coalesced droplet detaches from the substrate successfully. In the whole coalescence process, the energy difference $\Delta E$ is defined as the difference in $E_i$ between the beginning and the end of stage II, that characterizes the strength of the reaction force from substrate to propel the upward motion and detachment of droplet, and the duration of coalescence $\Delta t$ is defined as the time from the beginning of stage I to the end of stage II. It is obvious that substrate-droplet interaction energy difference on the wedged ridge is larger than that on the pillar array, $\Delta E_1 > \Delta E_2$, indicating that the wedged ridge could provide a stronger upward reaction force. Moreover, the coalescence on the wedged ridge is a shorter duration event, $\Delta t_1 < \Delta t_2$, thus the available energy is less dissipated throughout the coalescence process. Combining these two factors, it follows that the wedged ridge could effectively enhance the droplet jumping compared to the conventional pillar array.

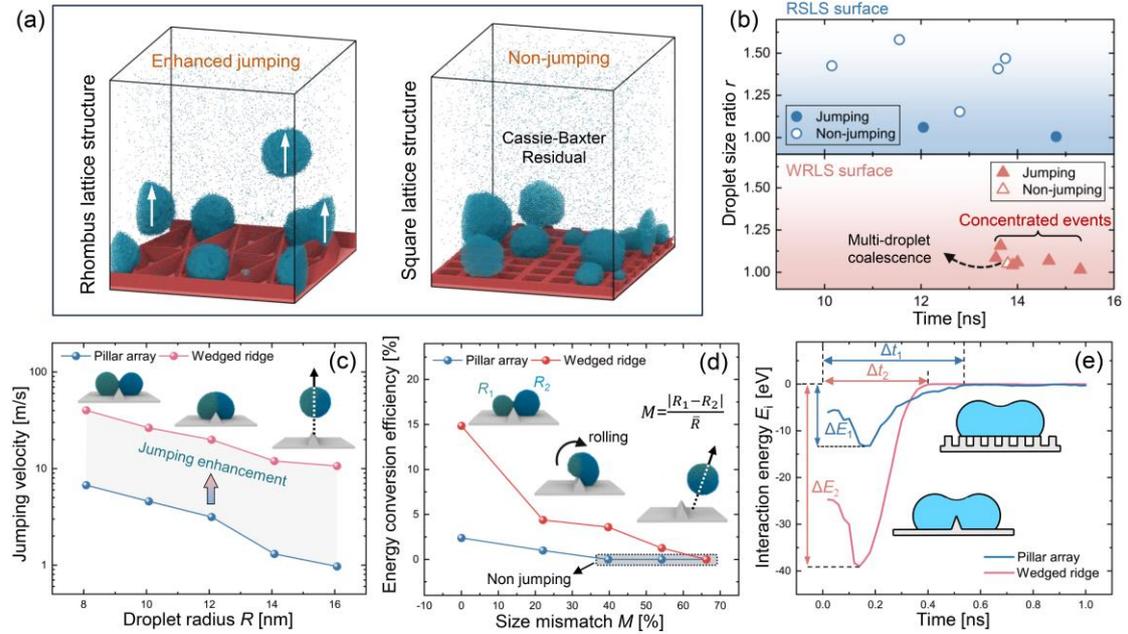

Figure 5. Enhanced coalescence-induced jumping: (a) snapshots of the out-of-plane removal of condensed droplets by coalescence on the WRLS surface and RSLS surface. Temporal Comparison for the droplet coalescence events between the WRLS surface and the RSLS surface: (b) the droplet size ratio $r$ and outcome of each coalescence event on WRLS surface and the RSLS surface, $r$ denotes the size ratio of larger droplet to smaller droplet before coalescence. Jumping enhancement of coalescing droplet by the wedge-shaped ridge of WRLS surface: (c) variations of jumping velocity with droplet radius, and insets are time-series snapshots showing the coalescence between equal sized droplets on the wedge-shaped ridge; (d) variations of energy conversion efficiency with size mismatch of coalescing droplet, insets are time-series snapshots showing the coalescence between two unequal sized droplets on the wedge-shaped ridge. Mechanism for the enhancement of coalescence-induced jumping: (e) variations of droplet-substrate interaction energy with time during coalescence.

*2.6. Periodical and sustainable refreshment*

By combining the spatially regular nucleation, bi-directional self-transport and enhanced coalescence-induced jumping of condensed droplets, the WRLS surface exhibits a sustainable self-renewal ability. Figure 6a shows the morphological evolutions and dynamic behaviors of condensates over its entire lifecycle on WRLS surface. It can be seen that the surface refreshment is closely associated with the droplet lifecycle, and thus presents a periodic pattern. Moreover, although a few residual droplets remain on the surface at the end of the first refresh cycle, they could be removed soon via coalescence-induced jumping in the subsequent refresh cycles, enabling a persistent control of droplet size. On the contrary, because of the inefficient droplet removal or shedding, the condensates accumulate on the RSLS surface and cover a large area gradually, that impedes the surface refreshment and thus induces the failure of jumping droplet condensation eventually, as shown in Figure S19 of Supplementary Information.

The persistent surface renewal was quantitatively analyzed by tracking the evolution of an isolated droplet within the unit cell of rhombus lattice, and Figure 6b correspondingly records the temporal variation of the droplet volume, which exhibit a stable periodicity. For each refresh cycle, the condensed droplet inflates first during the natural growth stage marked by the orange shaded area, and then suddenly detach from the surface during the coalescence stage marked by the blue shaded area, that leaves the solid surface exposed to the water vapor again and starts the next refresh cycle.

Moreover, the refresh period $\Delta t$, denoting the time span between two refresh cycles, is nearly constant with an average value of $\Delta t \approx 16.3$ ns. It follows that the condensed droplets are effectively removed in a sustainable manner, which shows great potential for eliminating the surface flooding caused by the condensate accumulation and further enabling an efficient jumping-droplet condensation with the long-term durability.

The overall renewal capability of surface was quantitatively assessed by considering the control on the residual droplet size. We separately extracted the largest residual droplet residing on the WRLS surface and the RSLS surface to calculate the equivalent diameters, whose variations with condensation time are shown in Figure 6c. On the RSLS surface, the residual droplet accumulates ceaselessly throughout the entire condensation without effective jumping and shedding from surface, and thus its size increase uncontrollably by natural growth and mutual droplet coalescence. Whereas on the WRLS surface, shortly after undergoing the early growth stage, the residual droplet size turns to fluctuates periodically aligning with the refresh cycles, and always remains lower than a diameter of 9.1 nm after long-term condensation, indicating the residual droplet size is controlled by the surface refreshment sustainably.

Because of the sustainable surface self-renewal, the condensates are removed timely and thus the exposed surface can directly contact with the heated steam over and over again, that would reduce the thermal resistance and further enhance the heat transfer performance in theory, therefore, we calculated the heat transfer coefficient (HTC) to quantitively verify this speculation. Based on the kinetic energy and potential energy of water molecules, the heat $Q$ released by the detached condensate droplets was first obtained by calculating the difference in total energy from the gas to liquid phase[35], then the heat transfer coefficient $h$ was calculated by dividing the released heat $Q$ by surface projected area $A$, subcooling $\Delta T$ and condensation time $t$, namely, $h = Q/At\Delta T$, calculation details are described in Section S12 of Supplementary Information. Figures 6c and 6d compare the time history of $h$ between RSLS surface and WRLS surface. For the RSLS surface, the HTC rises rapidly in the initiation stage where the uncovered surface is exposed directly to the water vapor. However, it soon begins to decline and gradually stabilizes at a value of $h = 0.68 \text{ kW}/(\text{m}^2 \cdot \text{K})$ because of the accumulation of condensates, that cover the surface, introduce the thermal resistance and thus inhibit the heat exchange between cooling surface and high-temperature water vapor. This heat transfer deterioration indicates the conventional nanopillar structures are unqualified for suppressing the surface flooding that induce the failure of jumping-droplet condensation. On the contrary, the HTC of WRLS surface increases stepwise along with droplet jumping detachments and gradually slows down to an approximate constant value when the phase transition process tends to be steady. Based on the exponential fitting data, the stabilized value of HTC was calculated to be $h = 244.81 \text{ kW}/(\text{m}^2 \cdot \text{K})$. This extremely high heat transfer performance is attributed to the nanometric droplet jumping, and it does not appear to degrade during the long-term condensation process, because the rhombus lattice structures can remove condensates persistently and further prevent the surface from flooding. Given all that, both the efficiency and durability of jumping-droplet condensation are improved by the proposed WRLS surface.

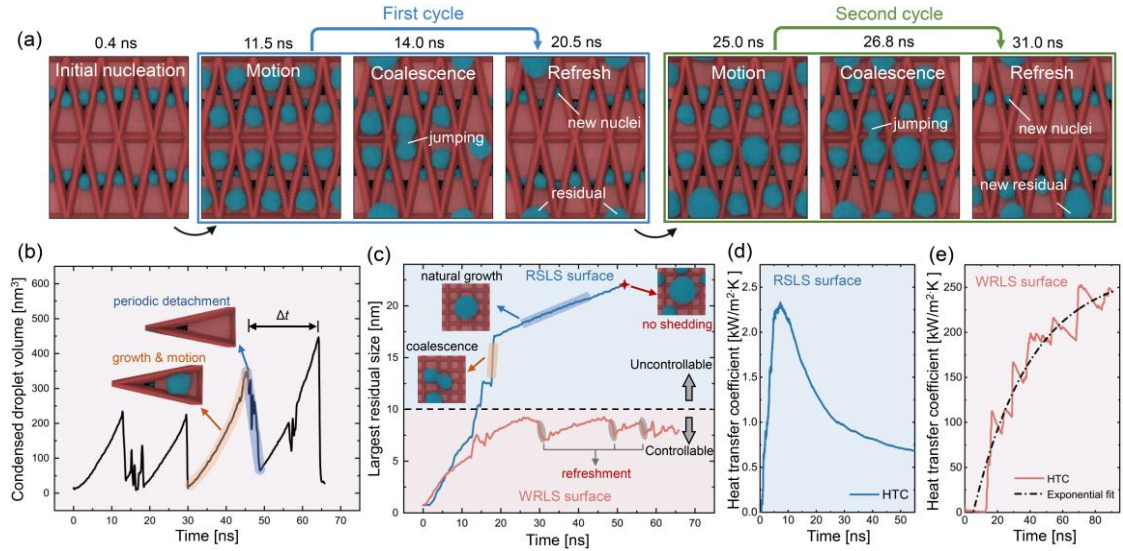

Figure 6. Periodical and sustainable refreshment: (a) time series snapshots in the top view shows the WRLS surface refreshment cycles involving persistent nucleation, migration, coalescence and removal of droplets; (b) temporal variation of the condensed droplets volume inside a unit cell of rhombus lattice of the WRLS surface. Comparison for the residual droplet size control capability: (c) time evolutions for the diameter of largest residual droplet on the WRLS surface (red line) and the RSLS surface (blue line), where the gradual increase represents the natural growth of isolated droplet, the abrupt decrease represents a successful detachment/refreshment induced by droplet coalescence while the abrupt increase denotes a droplet coalescence without jumping/refreshment. Comparison for the heat transfer performance between the WRLS surface and the RSLS surface: (d) temporal variation of the heat transfer coefficient for the RSLS surface; (e) temporal variation of the heat transfer coefficient for the WRLS surface.

## 3. Conclusions

We proposed a wedge-walled rhombus lattice structure (WRLS) surface based on the principles of biomimetics, and harness its multidimensional and persistent droplet rectification effects to realize a programmable jumping-droplet condensation with an excellent heat transfer efficiency and durability. Through MD simulations and theory analysis, the morphology evolutions and self-propelled motions of nanoscale condensates over the entire lifetime were studied systematically, and it is found that the effective and sustainable droplet detachment is implemented by the ingenious collaboration of regular nucleation, bi-directional self-migration and enhanced coalescence-induced jumping. Due to the lower nucleation energy barrier caused by the wetting contrast pattern and the small opening angle of lattice, the liquid embryos preferentially form at the converging sides of lattice structures and present a spatially regular and uniformly sized nucleation. In the subsequent growth stage, the inflating droplets are squeezed under the shape restriction of lattice frame, the resultant asymmetric Laplace pressure impel them to move toward the central diverging sides, and this spontaneous in-plane droplet rectification follows the principle of minimum energy. Moreover, the force-based model and simulations consistently suggest that the directional transport velocity of droplet rises as the opening angle of lattice decreases. At last, the droplets merge with each other and detach from surface efficiently with the assistance of intermediate wedged ridge, namely the out-of-plane droplet rectification. Owing to the strong upward reaction force and the short duration of coalescence, both the energy conversion efficiency and the size mismatch of coalescence-induced jumping are enhanced by the wedged ridge. By combining the above-mentioned unique properties, the condensed droplets are sustainably removed during the long-term condensation, that refreshes the surface frequently and thus eliminates the surface flooding. As a result, the heat transfer performance and durability of jumping-droplet condensation are simultaneously improved by the WRLS surface.

In summary, the droplets could actively coalesce with neighbors and further jump off surface by means of the persistent rectifications, this novel strategy enables the condensates to be removed frequently, greatly prevent the surface from flooding and further achieve a durable and highly-efficiency jumping-droplet condensation. The findings of this work provide new insights into the heat transfer enhancement for long-term applications while also promoting droplet manipulation techniques and self-cleaning surface technology.

## 4. Methods

All the MD simulations were performed using the open-source code LAMMPS[49]. The solid substrate is constructed by copper atoms (Cu) with a lattice constant of 3.61 Å, and a coarse-grained (CG) model[50], where each CG water bead (W) represents four water molecules, was adopted to constitute the vapor or droplet based on the corresponding density at a given temperature. Cu-Cu and W-W interactions were respectively described by the embedded atom model (EAM) potential[51] and Morse potential[50]. Lennard-Jones (LJ) 12-6 potential was employed to describe Cu-W interactions, and the energy parameter $\varepsilon_{Cu-W}$ was adjusted to achieve the desired contact angle. The detailed forcefields information are shown in Section S1 of Supplementary Information, and the equilibrium contact angles for the hydrophobic structures and hydrophilic patterns studied in this work are calculated to be 126.5° and 87.8° (please see Section S2 in Supplementary Information).

*4.1. Condensation*

All the condensation simulations involve two stages, we first conducted a pre-equilibration after the model energy was minimized, the entire system was integrated in an NVT ensemble with a Nose-Hoover thermostat at 500 K, to obtain the stable saturated vapor state, the variations of vapor temperature and pressure in the pre-equilibration stage are shown in Figure S5b in Supplementary Information. In the following main simulation stage, the substrate was switched to act as a cold source at 300 K, extracting heat and initiating condensation. To supply fresh vapor continuously and establish the saturated state of vapor domain throughout the condensation process, the GCMD (Grand Canonical Monte Carlo and molecular dynamics) region serving as a vapor/heat source at 500 K was applied at the top quarter of simulation domain[52]. The upper GCMD region and the remaining parts were integrated in grand canonical ($\mu$VT) ensemble and microcanonical (NVE) ensemble, respectively. Furthermore, another region below the GCMD area was defined to recognize the condensed droplets jumping from surface and remove them from vapor space, and its distance from surface was set to be larger than the cut-off radius of Cu-W interactions (Figure S5a in Supplementary

Information). For detailed simulation parameters and setup, please refer to Sections S1-S3 of Supplementary Information.

*4.2. Directional transport*

There are two types of MD simulations of directional droplet transport in this work, namely, the spontaneous migration of a growing droplet and the directional transport of a constant-sized droplet, on the identical unit cell of WRLS surface. For the former case, the simulation setups are fully consistent with that of the above-mentioned condensation simulations. For the latter case, a droplet was artificially constructed with its base was initially embedded in the converging end of the rhombus lattice. Similarly, this directional transport simulation involves two stages. A pre-equilibration was fist conducted after model energy minimization, the entire system was integrated in an NVT ensemble at 300 K to obtain an equilibrium wetting state, and in this stage, a spring force was applied to constrain the droplet and prevents its premature self-transport. Then, the droplet was switched to a microcanonical (NVE) ensemble, and the constraint was removed to initiate the directional transport. For detailed simulation parameters and setup, please refer to Sections S8 and S9 of Supplementary Information.

*4.3. Coalescence*

For the coalescence between two droplets on surface, the MD simulations were implemented by completing the following steps. As shown in Figure S16 in Supplementary Information, we first constructed two subsystems, each containing the left or right half of the surface and a single droplet residing atop the surface, these subsystems were separately integrated in an NVT ensemble at 300 K to obtain their corresponding equilibrium wetting states. Then we combined them to form a main system, and the coalescence-induced jumping behavior was naturally triggered by the intermolecular interactions between two droplets without any external disturbance. For detailed simulation parameters and setup, please refer to Section S10 of Supplementary Information.

**Acknowledgement**

We acknowledge the financial support from the National Natural Science Foundation of China (Nos. 52306088, 52076088, 52176078 and 52276067). This work was also supported by the Natural Science Foundation of Jiangsu Province (No. BK20230533) and the China Postdoctoral Science Foundation (No. 2023M731889).

**Author contributions**

S. G. conceived the research, performed the MD simulations and wrote the original manuscript. S. G., J. Q. and D. W. analyzed the simulation and theoretical data. Z. L. and W. G. supervised the research, provided funding and revised the manuscript.

**Competing interests**

The authors declare no competing financial interests.